# Unusual acceleration and size effects in grain boundary migration with shear coupling

Liang Yang[a], Xinyuan Song[b], Tingting Yu[c], Dahai Liu[a,*], Chuang Deng[b,*]

[a] School of Aeronautical Manufacturing Engineering, Nanchang Hangkong University, Nanchang 330063, China

[b] Department of Mechanical Engineering, University of Manitoba, Winnipeg, MB R3T 2N2, Canada

[c] School of Aviation and Mechanical Engineering, Changzhou Institute of Technology, Changzhou, Jiangsu 213032, China.

* Corresponding author: dhliu@nchu.edu.cn(D. Liu), Chuang.Deng@umanitoba.ca (C. Deng)

**Graphical abstract**

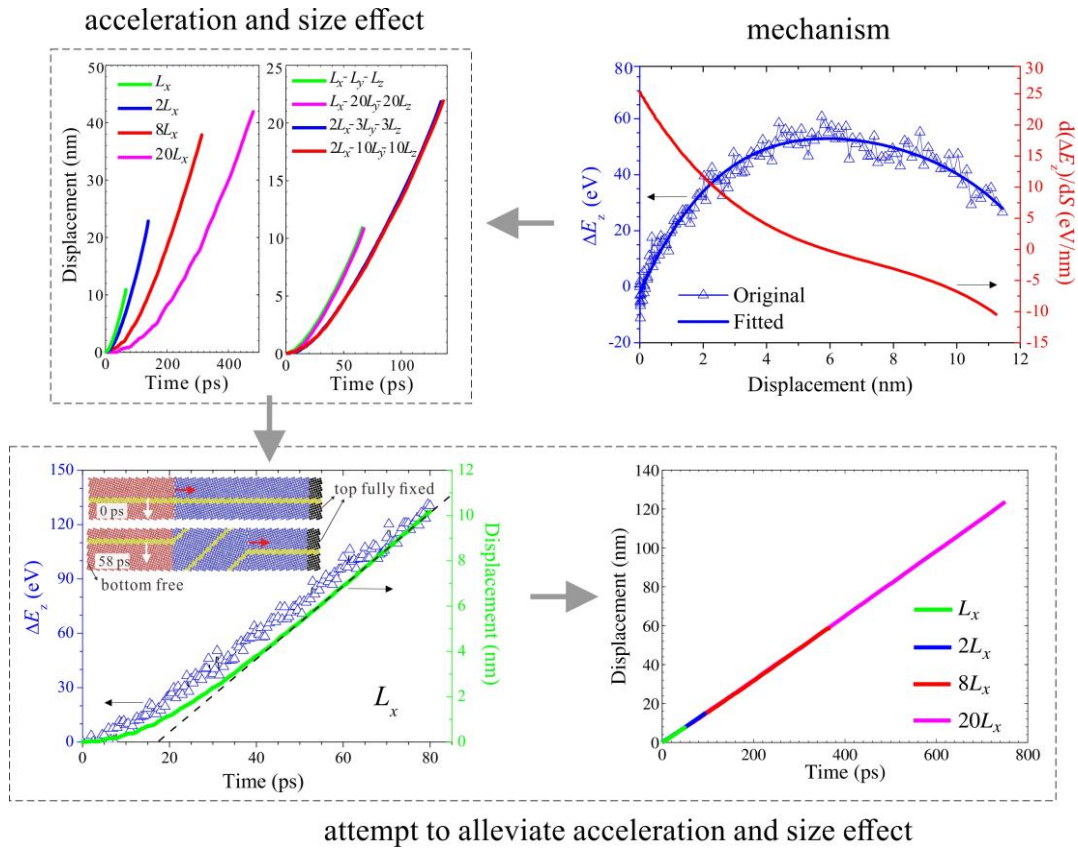

**Abstract**

Grain boundary (GB) migration plays a crucial role in the thermal and mechanical responses of polycrystalline materials, particularly in ultrafine-grained and nano-grained materials exhibiting grain size-dependent properties. This study investigates the migration behaviors of a set of GBs in Ni through atomistic simulations, employing synthetic driving forces and shear stress. Surprisingly, the displacements of some shear-coupling GBs do not follow the widely assumed linear or approximately linear relation with time; instead, they exhibit a noticeable acceleration tendency. Furthermore, as the bicrystal size perpendicular to the GB plane increases, the boundary velocity significantly decreases. These observations are independent of the magnitude and type of driving force but are closely linked to temperature, unique to shear-coupling GBs that display a rise in the kinetic energy component along the shear direction. By adopting a specific boundary condition, the acceleration in migration and size effect can be largely alleviated. However, the continuous rise in kinetic energy persists, leading to the true driving force for GB migration being lower than the applied value. To address this, we propose a technique to extract the true driving force based on a quantitative analysis of the work-energy relation in the bicrystal system. The calculated true mobility reveals that the recently proposed mobility tensor may not be symmetric at relatively large driving forces. These discoveries advance our understanding of GB migration and offer a scheme to extract the true mobility, crucial for meso- and continuum-scale simulations of GB migration-related phenomena such as crack propagation, recrystallization, and grain growth.

*Keywords*: grain boundary migration; shear-coupling; size effect; mobility; atomistic simulation.

## 1. Introduction

Grain boundary (GB) is crucial to a variety of behaviors (e.g., grain growth, recrystallization, and plastic deformation) and mechanical properties (e.g., strength and ductility) in polycrystalline materials [1]. One solid evidence is that nano-grained materials will exhibit significantly increased strength/hardness, improved toughness, and enhanced fatigue resistance in comparison with conventional coarse-grained materials, due to much higher volume fraction of GBs that reduces the grain size to the nanometer regime [2]. Meanwhile, the key role of GBs played in the deformation mechanism for nano-grained materials is no longer as barriers to slip transmission, as in conventional polycrystalline materials, but as the primary facilitators for plastic deformation, i.e., GB-mediated deformation processes (e.g., GB migration, sliding and grain rotation) [2-4]. Nano-grained materials nevertheless are inclined to be structurally unstable under external thermal or mechanical impetus, which can force the GB migration that has been recognized as a dominant scheme for softening or deviation from the Hall-Petch relationship [3,5,6]. To maintain the superior size-dependent properties of nanocrystalline metals, tremendous efforts have been paid to improve the microstructural stability through suppressing GB migration, which can be realized by adjusting GB structure, chemistry, temperature and so on [5,7,8]. Evidently, a comprehensive understanding of GB migration is of critical importance in probing and optimizing GB-related properties, especially for ultrafine-grained and nano-grained materials.

Up to now abundant and deep insights have been gained into the characters and underlying mechanisms of GB migration based on theoretical and experimental investigations [9-18]. Thereinto, dramatic attentions were paid to GB mobility $M$, which can be defined as the coefficient relating to migration velocity $v$ and driving force $P$ (i.e., $M = v/P$) and is considered an intrinsic GB property (i.e., only depending on material parameters, temperature, and boundary crystallography) [17-19]. Nevertheless, computational studies [20-23] have revealed that the magnitude of driving force can leave significant influences on mobility, owing to the force-induced variation of boundary structure and/or migration mechanism. For example, Deng and Schuh [20] found that for both symmetrical and inclined Ni $\Sigma 5 \langle 100 \rangle$ tilt GBs, their mobilities agree well with the intrinsic values obtained by the thermal fluctuation method [24] only when the applied driving force is sufficiently low; increasing the driving force will lead to diffusive-to-ballistic transition in the migration mechanism and enlarge the discrepancy between the extracted and intrinsic mobility values. Moreover, for shear-coupling migration GBs (i.e., simultaneous translation in GB plane during the migration along the boundary normal direction), Han and coworkers [14,25,26] demonstrated that both GB mobility and shear-coupling factor (ratio of GB sliding and migration rates) do not only strongly depend on the magnitude but also the source of driving force (stress or a jump in chemical potential across the boundary). They further revealed that the mobility traditionally defined as a scalar should be a symmetrical second-rank tensor [26]. The tensor components can be extracted by applying driving forces in the directions perpendicular and tangent to the boundary plane, respectively [26].

In addition to the driving force, GB motion may also strongly depend on the size of simulation cell. Zhou et al. [27] reported that the mobility of a Ni ⟨100⟩ tilt GB decreased monotonically with decreasing the cell thickness (the size along the tilt axis), due to the interference between the free surface and the collective rearrangement of atoms during boundary motion driven by an external stress. A similar size-dependency of mobility was also observed in Ref. [28]. Race et al. [29] revealed that the boundary area of a flat ⟨111⟩ tilt GB should reach the meso-scale or a large-enough value to yield a converged migration velocity under the synthetic driving force (SDF). Meanwhile, simulations for stress-driven [30,31] and SDF-driven [32] migration further discovered that the energy barrier for disconnection nucleation or the driving force for GB migration would converge when the boundary area was large enough for shear-coupling GBs. The energy barrier was also found to firstly decrease and then keep steady with the increase of cell size in GB normal direction for 53.1° Σ5 ⟨100⟩ tilt GB [32]. Existing studies concerning the size effect on GB migration overall reached a consensus that the system size should be large enough to yield physically reliable results and conclusions. This agrees well with the general understanding related to the size effect in modeling and simulation.

Although GB migration has been reported to suffer influences from various factors (e.g., crystallography [18,33], temperature [20,26], driving force [17,21], pressure [35,36] and impurity [37,38]), the mobility values extracted from $M = v/P$ in these studies were all based on a basic premise that the boundary velocity will maintain constant or approximately constant during the whole migration process under a fixed driving force, i.e., the boundary displacement exhibiting a linear or approximately linear relation with the migration time. This character has been widely observed in existing research concerning GB migration (e.g., Refs. [10,39-42]). Nevertheless, in this study, we found that velocities of some GBs did not keep constant during migration but exhibited an unusual acceleration feature, i.e., the velocity varying significantly with GB relative position in the simulation cell. This signifies a strong dependency of migration on the cell size in the direction perpendicular to GB plane, which has not yet been discovered before.

The first effort of the present work was therefore to investigate the underlying mechanisms for the acceleration in GB migration and effects of model size in GB normal direction on migration, based on atomistic simulations of several GBs driven by the external stress and SDF. After the corresponding conditions and mechanisms for these two phenomena were clarified, attention was paid to effectively alleviate the size effect and to extract the true driving force and mobility in the presence of acceleration. The present study will enhance our understanding of GB migration-mediated and grain size-dependent behaviors in polycrystalline materials. Additionally, it offers a method for extracting the true mobility, a fundamental property of GB, which is essential for GB-related simulations at meso- or continuum scales, such as crack propagation, recrystallization, and grain growth.

## 2. Methodology

In this study, the acceleration in GB migration, size effect and other related contents were investigated based on atomistic simulations of several GBs in Ni. First, we simulated the migration of Ni Σ5 ⟨100⟩{310}, Σ29 ⟨100⟩ {10 4 0} and Σ55 ⟨211⟩ {952} symmetrical tilt GBs, which correspond to P1, P148 and P233 GBs in the 388 GBs dataset constructed by Olmsted et al. [18], to reveal the phenomena of acceleration and dependency of migration on the cell size in the direction perpendicular to GB plane. These simulations were performed at 500 K and an SDF of 0.06 eV when increasing the normal cell size for each GB. Additional simulations were carried out at lower SDF (i.e., 0.025 eV for P1 and 0.003–0.006 eV for P233) and under an external shear stress $\tau_{ext}$ parallel to the z direction in Fig. 1(a) ($\tau_{ext}$ = 250 MPa for P1 and $\tau_{ext}$ = 50 MPa for P233) to test whether the above phenomena are affected by the magnitude and type of driving force, respectively. Second, we simulated the migration behaviors for Ni Σ3 ⟨110⟩ twist GB (i.e., P5) at 500 – 1000 K and 0.006 – 0.06 eV, Σ21 ⟨210⟩ general GB (i.e., P81) at 1000 K and 0.001 – 0.06 eV, P1 at 1000 K and 0.06 – 0.003 eV. These simulations were aimed to explore the physical causes for the acceleration and size-dependency from the aspects of shear coupling and work-energy relation in the bicrystal system. Third, we chose the P1 GB as the representative example to attempt approaches for inhibiting or alleviating the size effect by performing simulations adopting various boundary conditions (BCs) and/or manipulating the internal stress in the system. The corresponding results were compared with those reported in existing studies whenever possible. Finally, we still chose P1 as the example to discuss how to extract the true driving force for GBs exhibiting acceleration when the true driving force was not equal to the value nominally applied through the SDF method or the external shear stress.

All simulations stated above were performed using the Large-scale Atomic/Molecular Massively Parallel Simulator (LAMMPS) software package [43] with the embedded atom method potential developed for Ni [44]. As shown in Fig. 1(a), a bicrystal simulation cell was used to construct a flat GB, which was in *y-z* plane with *y* and z directions being periodic. Nevertheless, the boundary conditions in *x* direction (parallel to the boundary normal direction) might be quite different, depending on the simulation tests. For the above first and second groups of simulations, in order to avoid translation of the whole bicrystal in GB normal direction, a slab of atoms (1nm thickness) near the bottom surface were partially fixed (i.e., only the velocity and force components along the *x* direction being set as zero) while the top surface was set free. For the third group of simulations, the boundary conditions in *x* direction might be periodic, fully fixed, free, or one surface free while the other fixed. When exploring the size effect on boundary migration, the cell size in the GB normal direction (i.e., the grain size) or in the boundary plane (i.e., the boundary area) was varied accordingly. Table 1 summarizes the boundary conditions applied in this study and the corresponding results simulated under individual conditions. The minimum cell size for each above GB was the same as those constructed by Olmsted in Ref. [18]. For example, the minimum cell size for P1 was $L_x$ = 18.3 nm, $L_y$ = 3.2 nm, and $L_z$ = 3.3 nm.

When the bicrystal system of each GB was constructed, its energy was minimized at 0 K following the scheme introduced in Ref. [18]. Subsequently, the system was elevated to and sufficiently relaxed at test temperatures (500 K or 1000 K) under the isothermal-isobaric ensemble (NPT) for about 0.15–7 ns (depending on the temperature, GB, and cell size) with a default time step of 5 fs. After the system was fully equilibrated, the boundary was driven to migrate under a jump in chemical potential across the boundary or a shear stress. Thereinto, the former was realized by the CROP-SDF method [21] while the latter by applying a shear force to individual atoms near the bottom surface (1 nm thickness in *x* direction) in Fig. 1(a). The GB displacement under the SDF and stress was computed by tracking the overall change of the potential energy artificially added to the bicrystal system and by tracking atoms with the centro-symmetry parameters close to the maximum value in the system, respectively. For the above four groups of simulations, the NPT ensemble was also used during the boundary migration for each case, if not otherwise specified, to control the internal normal stress components as close to 0 GPa as possible. For some specific simulations, the internal shear stress along the shear direction also needed to be controlled at 0 GPa. The structure of bicrystal model, if needed, was visualized by Ovito package [45]. Note that some results concerning this study were presented in the Supplementary Material.

**Table 1** Types of boundary conditions applied along the GB normal direction (i.e., *x* direction) in this study and the corresponding simulation results under each boundary condition. The other two directions in the GB plane are both set to be periodic.

| Boundary conditions | Corresponding results |
| --- | --- |
| Set the bottom surface as partially fixed while the top surface as free | Figs. 1-4 and s1 |
| Set the surface as periodic | Fig. 5 |
| Set the top and bottom surfaces as fully fixed | Fig. 6 |
| Set the surface as periodic while controlling shear stress close to 0 GPa | Fig. 7 |
| Set the bottom surface as free while the top one as fully fixed | Figs. 8-10 and s2, Table 2 |

## 3. Results and discussion

3.1 *Acceleration in migration and size effect*

Fig. 1(b-e) represents the migration data of Σ5 ⟨100⟩ {310} tilt GB (i.e., P1 GB) when increasing the cell size along different directions while under a constant external driving force. Two features can be readily observed. First, the boundary velocities (slope of displacement curve) under various cell sizes all gradually increase with the proceeding of boundary migration, suggesting a clear dependency of migration velocity on the relative GB position along the boundary normal direction in the simulation cell. In contrast, the boundary displacement was widely observed and commonly assumed

to exhibit a linear or approximately linear relation with the migration time (e.g., Refs. [10,39]). After a detailed survey of existing studies, the acceleration in migration was only found in the work by Coleman et al. [46], who simulated the migration of Ni Σ37 ⟨100⟩ symmetrical tilt by applying the synthetic driving force and shear strain at 300 and 400 K. Unfortunately, the focus in Ref. [46] was the atomic mechanisms of migration and no attention was paid to the acceleration.

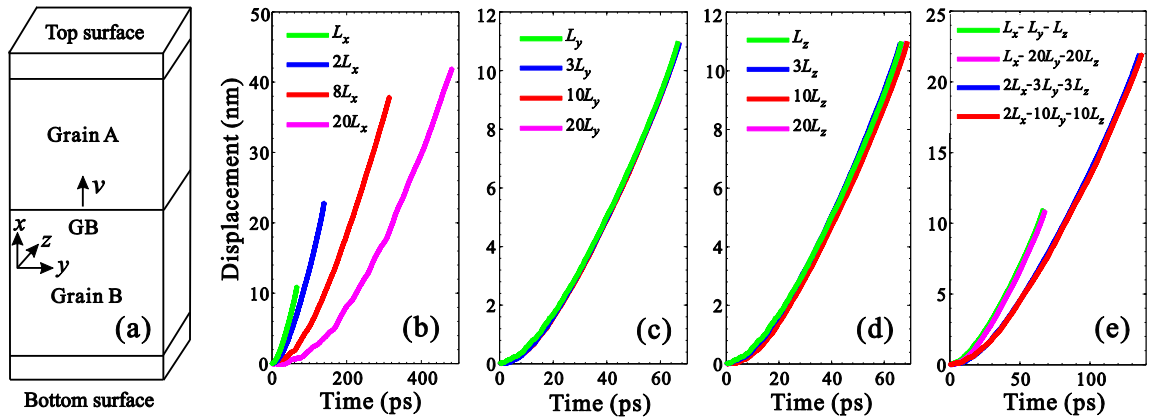

**Fig. 1** (a) Schematic of the bicrystal simulation cell. GB displacement *vs.* time for P1 GB simulated at 500 K and under a synthetic force of 0.06 eV, when increasing the cell size along (b) *x*, (c) *y*, (d) *z* and (e) all three directions. To avoid translation of the whole bicrystal along the *x* direction, the bottom surface was partially fixed (i.e., setting the velocity and force components along *x* for atoms near the surface as zero) while the top surface was set free.

Second, the velocity is independent of the cell size in GB plane (i.e., the boundary area or lateral cell size) (see Fig. 1c-e) but strongly and negatively related to the size in the boundary normal direction (*x* direction in this study) (see Fig. 1b and e). Nevertheless, the velocity of flat boundary has been previously reported to show a strong and complex dependence on the boundary area [29]. In addition to the velocity, the boundary area has also been reported to cause significant influence on GB mobility [27,28] and the energy barrier of disconnection nucleation for GB migration [30-32]. Moreover, these properties concerning GB migration exhibited a consistency in their dependency on the boundary area, i.e., the boundary area should be sufficiently large to yield a converged property value. The negative dependency of velocity on the cell size along the boundary normal (i.e., vertical cell size) here is partially similar to the trend regarding the threshold driving force of disconnection nucleation for Cu Σ5 ⟨100⟩ {210} tilt GB (i.e., P6 GB in Ref. [18]) at 10 K revealed by Deng and Deng [32], who found the threshold driving force, which in practice can be qualitatively regarded as the reverse of GB mobility [16], overall declines when increasing the vertical cell size. Therefore, the size effects regarding the boundary area are different between the present and existing studies, but a similarity appears in the dependency on the vertical cell size.

As a typical of low-period and high-angle CSL boundary, the migration behaviors of P1 GB have been widely studied through atomistic simulations [13,17,17,29,21,25,42,47-49], but why the above two features were not reported before? This can be attributed to multiple factors. First of all, there has been no research adopting various vertical cell sizes for this GB up to now, and accordingly no insight into the size effect was obtained. In studies adopting fixed vertical size [17,18,21,25,42,47], the acceleration might also exist, though the displacement-time data was not directly provided in these studies. Nevertheless, we deem that the acceleration might have been disregarded on the grounds that the main attentions and efforts were focusing on exploring the intended objectives of individual studies, as in our previous work [21,42]. In a recent work by Starikov et al. [50] focusing on the disordering complexion transition of GB, the acceleration was observed for Mo 22.6° [001](510) GB while there was not any discussion or analysis about it. Meanwhile, the disappearance of acceleration can also be attributed to the relatively high temperatures (e.g., 1000, 1200 and 1400 K) tested in Refs. [17,18,47] (see discussion in Section 3.2). In addition, the periodic boundary condition imposed along the boundary normal direction will prevent the presence of acceleration in work [20,48]. This boundary condition has been confirmed to inhibit the shear-coupling migration [29,49], which is a necessary but not a sufficient condition for acceleration migration (see following discussion concerning Figs. 2 and 3). When exploring the shear coupling migration of Cu P1 GB, Cahn et al. [13] directly presented a displacement-time data up to 1 nm, simulated by applying a constant shear strain 1 m/s at 800 K (see Fig. 6 in Ref. [13]). To our understanding, 1 nm data may not be sufficiently long to evidently illustrate the acceleration feature, in comparison with the displacement data in Fig. 1. Another displacement-time data (up to 6 nm) was provided by Schratt and Mohles [49], who simulated the migration of Ni P1 under 300–1000 K with free end boundary conditions and a synthetic force of 0.06 eV imposed through the ECO-SDF method. Nevertheless, the acceleration feature was still not observed in Ref. [49] though it shows up in our re-tests of simulations in Fig. 1 by using the ECO-SDF method. Therefore, the discrepancy concerning the acceleration should not be attributed to the different versions of SDF method (i.e., CROP-SDF [21] or ECO-SDF [49]) utilized in the present study and Ref. [49]. Since the temperature dependency of migration velocity and shear coupling factor $\beta$ in the range of 300-700 K in [49] also appears different from those previously reported [13,16,20,25], we deem that the discrepancy may be resulted from the difference in the metastable structures for P1 GB adopted for various studies.

To evaluate whether the acceleration in migration or the corresponding size effect is unique to P1 GB or not, Fig. 2 show the results simulated for some other GBs or under simulation settings different from Fig. 1. As shown in Fig. 2(a) and (b), the two features can also be observed for P148 and P233 GBs when adopting the same settings as for P1 in Fig. 1. Since the driving force applied in Figs. 1, 2(a) and (b) is a relatively high value (i.e., 0.06 eV ≈ 0.87 GPa) in comparison with experimentally applied values, we tested lower forces for P1 and P233 GBs. It can be seen from in Fig. 2(c) and (d) that these features still hold on for P1 GB at 0.025 eV (lower than $KT$ = 0.043 eV, $K$ Boltzmann constant and $T$ temperature) and for P233 GB at 0.003 eV which approaches typical experimental values. Note that lower forces have also been tried for P1 but failed to yield continuous boundary migration, agreeing with the threshold driving force of boundary migration determined for this GB at 500 K by Yu et al. [16]. It is important to note from Fig. 2(e) and (f) that the acceleration and size effect still show up for P1 and P233 GBs when applying the external shear stress to drive the GB migration. Therefore, while the shear coupling mode may be strongly influenced by both the magnitude and type of the driving force [25], the acceleration and size effect does not exhibit such dependency.

The above analysis suggests that the acceleration in migration and negative dependency on the vertical cell size are relatively common features for force-driven GB migration. The following content will further reveal that the latter feature is resulted from the former one. These two features extend our current understandings of size-effect on GB migration, which almost all focused on the size in the boundary plane [27-31]. They also remind us that attentions should be taken for GBs exhibiting such features when extracting the boundary velocity or mobility under a constant driving force, during which the migration displacement and time were almost always assumed to keep a linear relation.

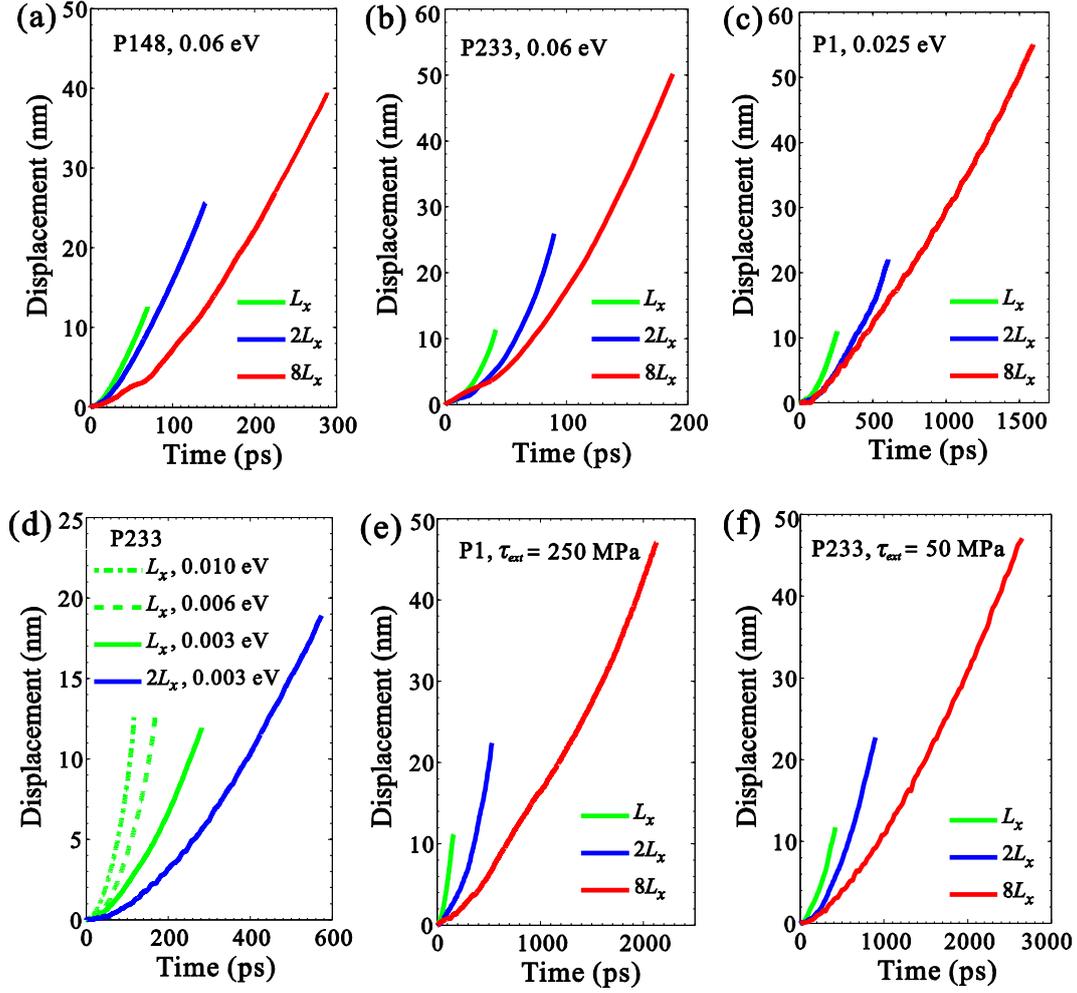

**Fig. 2** Other results simulated at 500 K supporting the acceleration in migration and size effect. Displacement data in (a-d) were all simulated under the applied synthetic force while (e, f) under the external shear stress. The shear stress was applied to a slab of atoms (1 nm thickness) near the top surface, as illustrated in Fig. 1, while a slab of atoms near the bottom surface was set as a rigid body.

### 3.2 *Underlying mechanism for acceleration and size effects*

The acceleration in boundary migration and vertical size effect have been revealed in the above section, then for what types of GB or under what kinds of condition that such phenomenon will occur? A preliminary analysis of the three GBs tested in Figs. 1 and 2 indicates that they are all shear-coupling migration GBs and with $\beta > 0.5$. Fig. 3(a) chooses P5 (Ni Σ3 ⟨110⟩ twist) GB as an example to show the displacement-time ($S$-$t$) data and size-dependency of GBs without shear-coupling. It can be seen that the displacement is linearly related to the migration time and the corresponding velocities are the same under different vertical sizes, irrespective of temperature and magnitude of driving force. These results seemingly indicate that only shear-coupling GBs will exhibit acceleration migration and negative size-dependency. To yield shear-coupling migration for such GBs, atoms in the bicrystal system should be accelerated up to the lateral shear velocity, and the longer time should be theoretically needed for the acceleration in the case of systems with larger vertical sizes (i.e., the overall migration velocity is lower for larger system). However, as shown in Fig. 3(b) for P81 GB, the linear *S-t* relation and constant *v* under different sizes can be observed also for shear-coupling GBs at various temperatures and driving forces. Furthermore, the acceleration migration and size effect observed at 500 K for P1 GB (Fig. 1) unexpectedly transfers into uniform migration when raising the temperature to 1000 K at which the shear coupling still exists, regardless of the magnitude of driving force (see Fig. 3(c)). This kind of transition induced by the temperature also occurs for other shear-coupling GBs (see Fig. s1 in the Supplementary file). To our understanding, the transition can be attributed to the temperature-induced variation of disconnections mediated for GB migration, which has been widely observed [14,31]. In addition, the transition may be also attributed to the possible conversion of the thermal energy into the shear kinetic energy, which should be more readily at higher temperatures and then shortens or even call off the process of accelerating atoms up to the shear velocity. Based on these analyses, we conclude that the shear coupling is a necessary but not a sufficient condition for the acceleration in migration and therefore the size effect, which may suffer strong influence from the temperature.

To further explore the fundamental mechanisms for acceleration, Fig. 3(d) compares the relative variation of kinetic energy ($\Delta E_i$, $i = x$, $y$ or $z$) to the initial state for P1 GB at 500 and 1000 K, at which the boundary exhibits accelerated (Fig. 1(b)) and uniform (Fig. 3(c)) migration, respectively. At 1000 K, all three components of the kinetic energy remain almost unchanged (i.e., $\Delta E_x = \Delta E_y = \Delta E_z \approx 0$) with the proceeding of migration. In contrast, at 500 K, although $\Delta E_x$ and $\Delta E_y$ still remain unchanged, $\Delta E_z$ firstly increases and then decreases (the final $\Delta E_z$ is still much higher than zero). Note that the shear movement is parallel to *z* direction. The comparison suggests that the work ($W_{ext}$) done by the external driving force ($P_{ext}$) only contributes to shear-coupling migration at 1000 K, but to both shear-coupling migration and a rise in the shear kinetic energy ($\Delta E_z$) at 500 K. This difference reminds us that the accelerated and uniform migration can be qualitatively justified from the aspect of true driving force ($P_{true}$) for boundary migration, which can be influenced by $W_{ext}$ and $\Delta E_z$.

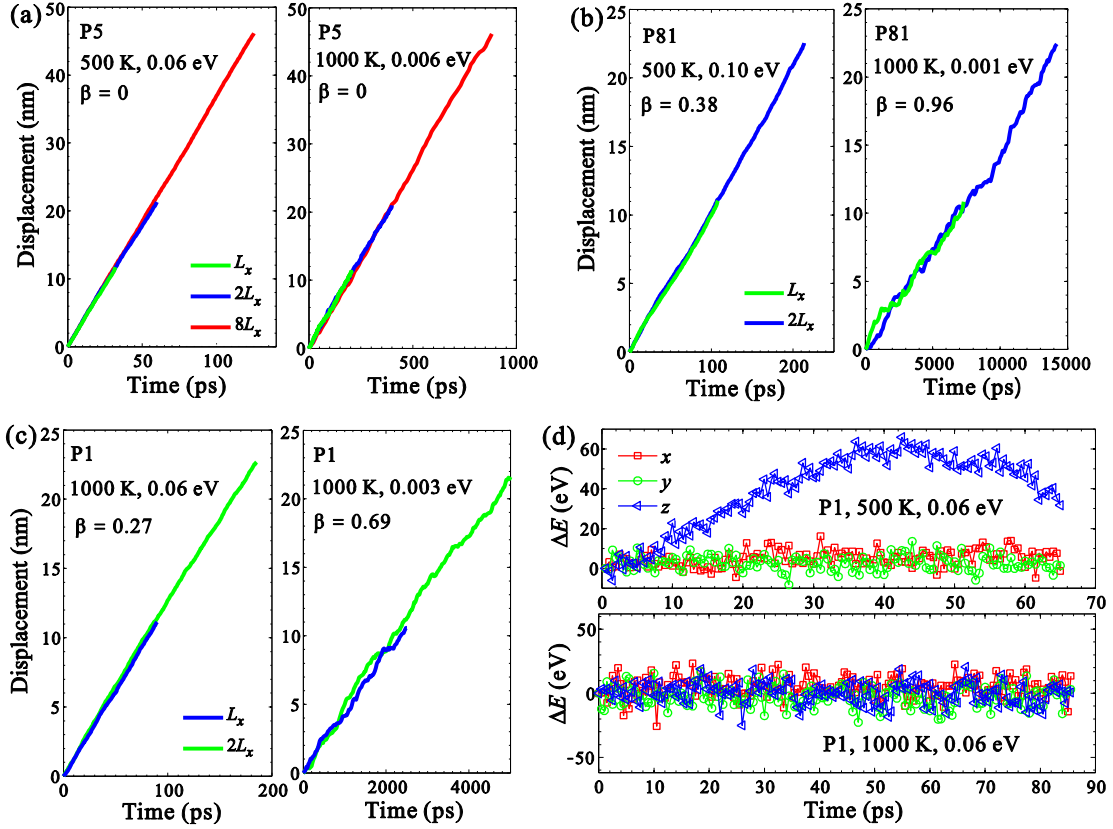

**Fig. 3** Examples of uniform migration for GBs with or without shear-coupling: (a) P5; (b) P81; (c) P1. (d) presents kinetic energy $\Delta E$ for P1 with cell size $L_x$, simulated at 500 K and 1000 K, respectively.

During the process of boundary migration, the work and kinetic energy for the bicrystal system meet the relation of $W_{ext} = W_{true} + \Delta E$. $W_{true}$ is the work done by $P_{true}$, i.e., $W_{true} = P_{true} \cdot S \cdot A_{GB}$, $S$ and $A_{GB}$ stand for the GB displacement and area, respectively. Considering $\Delta E_x = \Delta E_y \approx 0$ in the case of both accelerated and uniform migration, the relation can be given as $W_{ext} = W_{true} + \Delta E_z$. At one specific moment of migration, the work-energy relation can be further described as $P_{ext} \cdot dS = P_{true} \cdot dS + d(\Delta E_z)/A_{GB}$, and the instant true driving force is $P_{true} = P_{ext} - (d(\Delta E_z)/dS)/A_{GB}$, where $dP_{ext}$ is a fixed value for the SDF method. Therefore, $P_{true}$ will be constant when $\Delta E_z$ keeps unchanged (e.g., 1000 K at Fig. 3(d)), i.e., $P_{true} = P_{ext}$. In such case, the boundary will accordingly exhibit uniform migration (i.e., a linear $S$-$t$ relation) and thus consistent velocities when adopting distinct vertical sizes but the same $P_{ext}$ (e.g., Fig. 3(c)). Nevertheless, in the case of accelerated migration (i.e., $\Delta E_z \neq 0$, see Fig. 3(d)), $P_{true}$ depends on both $P_{ext}$ and $d(\Delta E_z)/dS$. From the $d(\Delta E_z)/dS$ vs. $S$ curve (the red curve) at 500

K for P1 GB shown in Fig. 4, we can observe that d($\Delta E_z$)/d$S$ continuously descends with the boundary migration, suggesting a continuous rise in $P_{true}$ and thus in migration velocity. Moreover, it is conceivable that when applying the same $P_{ext}$ to bicrystal systems with different vertical sizes (e.g., Fig. 1(b)), $P_{true}$ will be lower (i.e., lower velocities) for larger systems due to higher d($\Delta E_z$), which can be further attributed to more atoms involving shear movement for larger systems. When applying this interpretation to justify the size effect for systems with different sizes in the boundary plane (e.g., Fig. 1(c)), the contribution of GB area to $P_{true}$ must also be considered. In summary, the above analysis suggests that the acceleration in migration and negative dependency of velocity on the vertical size are unique to shear-coupling GBs exhibiting a rise in the kinetic energy component along the shear direction and can be justified from the aspect of true driving force based on the work-energy relation in the bicrystal system.

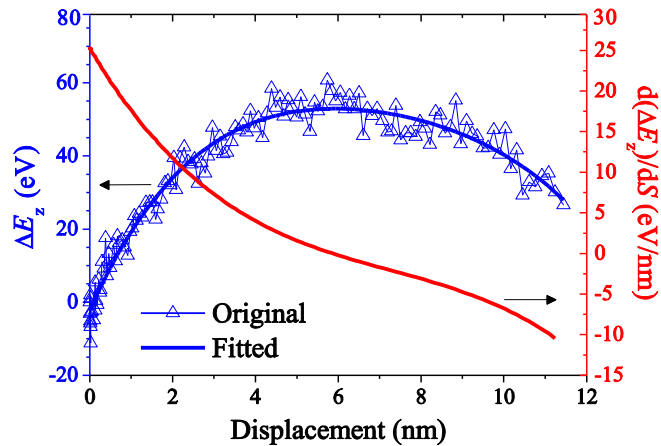

**Fig. 4** Variation of d($\Delta E_z$)/d$S$ with the boundary displacement for P1 GB, calculated based on the $\Delta E_z$-$S$ curve (blue curve) obtained by the least-square fitting of the original data at 500 K and 0.06 eV, given in Fig. 3(d).

3.3 *Attempts to alleviate acceleration*

Although the acceleration in migration has been demonstrated as a relatively common feature for force-driven migration of flat GBs in Section 3.1, it is undesired if the purpose is to compute a GB mobility by assuming $v = MP$. Then, is it possible to inhibit or alleviate the acceleration? For this purpose, we have carried out a series of simulations by manipulating the boundary conditions and internal stress in each bicrystal system (see Figs. 5-7).

Firstly, we tried to adopt periodic boundary condition in the GB normal direction ($x$ direction in Fig. 1(a)), which is one kind of boundary conditions widely used in previous studies [20,39,49]. It can be seen from Fig. 5(a) that the boundary displacements under various vertical sizes all nearly exhibit a linear relation with the time; the velocity nevertheless keeps increasing when enlarging the vertical size, in contrast to a negative size dependency of velocity in Fig. 1. Meanwhile, in comparison with 0.06 and 0.025 eV applied for the boundary condition adopted in Figs. 1 and 2c, much larger driving force (i.e., 0.15eV) must be applied to initiate the boundary movement for all cell sizes, suggesting a significant effect of boundary condition. Under the present boundary condition, the internal shear stress sharply increases with the initiation of GB migration, and then experiences a short descending and finally fluctuates around a very high value (see Fig. 5(b)), suggesting that the bicrystal system is far from reaching a steady-state. Furthermore, the shear stress is obviously lower under larger cell size. The higher velocity under larger size in Fig. 5(a) is therefore resulted from this tendency of shear stress, which can be attributed to more elastic energy released with the boundary migration due to larger space along the GB normal. As shown in Fig. 5(c), the periodic surface strongly inhibits the overall relative shear movement between two grains, as revealed in Refs. [20,49]), but only enables local shear movement which is more evident for larger cells. This precisely accounts for the linear $S$-$t$ relation under various sizes in Fig. 5(a), according to the discussion concerning mechanism for acceleration in Section 3.2. Evidently, the periodic boundary can effectively eliminate the acceleration but not the size effect by significantly constraining the shear movement and rendering the GB migration under an unsteady state.

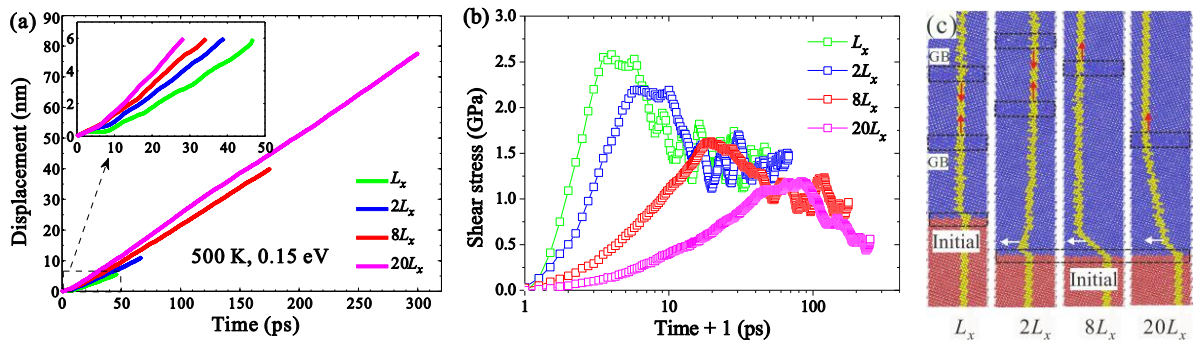

**Fig. 5** Migration results *vs.* vertical size for P1 GB, simulated at 500 K and 0.15 eV by adopting periodic boundary conditions along the GB normal direction: (a) displacement data; (b) shear stress; (c) snapshot of boundary migration. Red and white arrows in (c) illustrate the migration and shear directions, respectively.

Atoms are colored by atom type to visualize the shear-coupling migration using the Ovito software [45].

Secondly, we tried to set the top and bottom surfaces to be fully fixed. It can be seen from Fig. 6 that most of the results are similar to those under periodic boundary in Fig. 5, e.g., much larger driving force, linear $S$-$t$ and very high shear stress. Additionally, displacements are nearly independent of the system size (Fig. 6(a)), though the normal stress continues to rise with the boundary movement (see the example shown for cell size of $L_x$ in Fig. 6(b)). Evidently, the GB migration under the present boundary conditions is also under an unsteady state. The corresponding velocity is much lower than that at 0.06 eV in Fig. 1 while close to that at 0.15 eV and $L_x$ in Fig. 5. In consistency with the periodic boundary, the boundary condition of fixed ends also inhibits the global shear movement and enables only local shear. Fig. 6(c) shows that the local shear-coupling mode may change even switch under fixed ends, as already observed in Ref. [51].

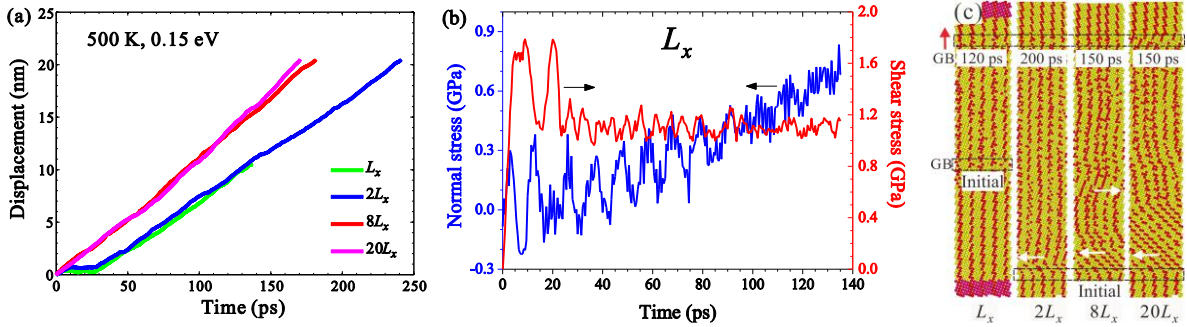

**Fig. 6** Migration results for P1 GB simulated at 500 K and 0.15 eV while setting the top and bottom surfaces to be fully fixed: (a) displacement data; (b) shear stress and normal stress under cell size $L_x$; (c) snapshot of boundary migration. The red and yellow atoms colored in (c) are aimed to visualize the local shear movement and shear-switching.

Thirdly, considering the impeding effect of the high internal shear stress on shear movement, we performed simulations that controlling the stress as close to 0 GPa as possible under periodic boundary (see Fig. 7). Although shear stresses are well controlled especially for larger systems (Fig. 7(b)), acceleration and size effect still exist (Fig. 7(a)). Moreover, the boundary stagnates long before reaching the other end, and the final displacement value under each size is nearly half of the feasibly maximum value (compare Figs. 5(a) and 7(a)). The inclined line displayed by the yellow atoms in Figs. 7(c) illustrates that the shear movement is inhomogeneous along the GB normal direction, and the top and bottom parts of the blue grain make shear along two opposite directions (see the white arrows). These results should be resulted from the cell inclination when controlling shear stress (Fig. 7(c)), which leads to the variation of crystallographic orientation and thus erroneous exertion of the orientation-dependent driving force for shear-coupling migration. We have also tried to control shear stress for fixed and free boundary conditions but as well obtained cell inclination and other results similar to those by using periodic boundary. Evidently, the above three attempts all failed to achieve our anticipated objectives of effectively alleviating the acceleration and size effect.

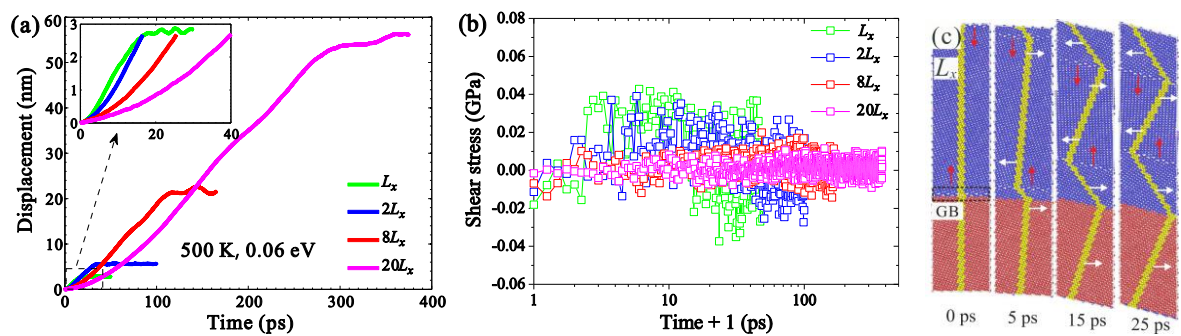

**Fig. 7** Migration results *vs.* vertical cell size for P1 GB, simulated at 500 K and 0.06 eV by adopting periodic boundary while controlling shear stress close to 0 GPa: (a) displacement data; (b) shear stress; (c) snapshot of boundary migration.

Finally, considering the inhibition of overall shear movement by periodic and fixed boundaries, we performed simulations adopting two free boundaries or setting one boundary as free while the other as fixed. It can be seen from Fig. 8(a) that the displacement data under two free boundaries and one free while another partially fixed are consistent and exhibiting an acceleration tendency (as observed in Fig. 1(b-e)), because the two grains across the GB are free to shear under these two boundary conditions. When the bottom surface is fully fixed (i.e., the grain near this surface is not allowed to shear), significant acceleration is observed throughout the whole migration process (the blue curve in Fig. 8(a)). Nevertheless, in the case of fully fixed top surface, acceleration is only significant at the early migration stage and gradually turns into uniform migration at the later stage (see the green curve and black dashed line in Fig. 8(a)), suggesting a gradual weakening of acceleration. The tendencies shown in the variation of migration velocity with the time under these boundary conditions can be more readily captured from the quantitative comparison of the instantaneous *v-t* data in Fig. s3 in the Supplementary file.

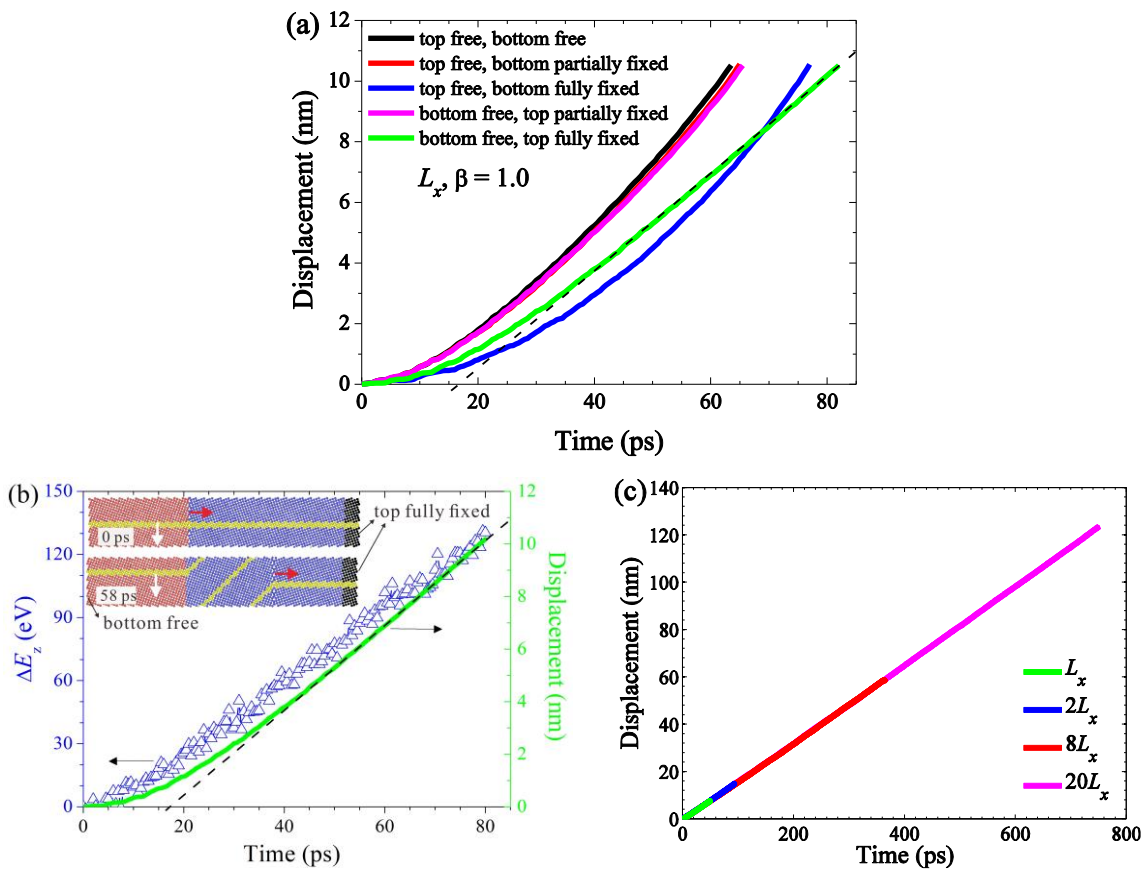

**Fig. 8** Attempts to alleviate acceleration by adopting one or two free boundaries for P1 GB with normal size $L_x$, simulated at 500 K and 0.06 eV. (a) Comparison of displacement data under different boundary conditions. (b) Variation of displacement, kinetic energy and snapshot with the time when setting the bottom surface as free while the top one as fully fixed. (c) Comparison of the linear *S-t* segments extracted from the complete displacement data under various cell sizes, simulated by adopting the same boundary condition as in (b). When the surface is partially (as in Fig. 1) or fully (setting all force and velocity components for atoms near the surface as zero) fixed, the grain near this surface can or can not make overall shear movement. The complete displacement data for (c) can be found in Fig. s2 in the Supplementary file.

The tendency in the case of fully fixed top surface can be justified from the similarity between the variations of $\Delta E_z$–*t* and *S-t* curves in Fig. 8(b) and from the relation $P_{true} = P_{ext} - (d(\Delta E_z)/dS)/A_{GB}$. Moreover, as illustrated by the inset snapshots in Fig. 8(b), the overall shear and coupling factor are not influenced by this boundary condition. Furthermore, Fig. 8(c) presents the comparison of the linear *S-t* segments extracted from the complete displacement data under various sizes. Interestingly, velocities are consistent for different cell sizes. Therefore, the size effect for shear-coupling GB can be considered as being eliminated if only focusing on the uniform migration stage. With these attempts, we may conclude that setting the top surface (i.e., in the forward direction of GB migration) of the bicrystal as fully fixed while the bottom surface (the backward direction of migration) as free is a relatively effective way to largely alleviate acceleration migration and thus size-dependency.

3.4 *Extraction of true driving force and mobility*

Although the accelerated migration and size effect for shear-coupling GBs can be effectively weaken by adopting one special boundary condition, the kinetic energy of the system $E_z$ still continues to rise during the boundary migration (see Fig. 8(b)), and thus the true driving force $P_{true}$ does not equal to the externally applied value $P_{ext}$ and depends on the variation of $E_z$ (see discussion in Section 3.2). Therefore, efforts should be paid to extract $P_{true}$ and the corresponding true mobility $M_{true}$. As already discussed in Section 3.2, $P_{true}$ can be determined based on a quantitative analysis of the work-energy relation in the bicrystal system. Meanwhile, considering the continuous rise of $\Delta E_z$ with boundary migration, we may not obtain a constant but a time dependent $P_{true}(t)$, and therefore the quantitative analysis should be carried out for individual steps of GB migration. The artificial energy added to the system by the SDF method in time interval d$t$ can be written as:

$$dE = \Delta e \cdot dn \qquad (1)$$

where $\Delta e$ denotes the maximum potential energy added to a single atom (e.g., 0.06 eV in Fig. 1) and d$n$ indicates the number of atoms whose corresponding crystallography changed as the boundary migrates. d$n$ can be calculated by d$n = (N/L_x) \cdot v(t) \cdot dt$. Here, $N$ and $v(t)$ represent the total number of atoms in the system and the instant migration velocity perpendicular to GB plane, respectively. However, due to the continuous rise of $E_z$, the actual energy to drive boundary movement is

$$dE' = dE - dE_z = dE - c_z \cdot dt \qquad (2)$$

where $c_z$ is variation rate of d$E_z$ with respect to time. We can thus deduce the true maximum energy imposed on per atom $\Delta e'(t)$ as:

$$\Delta e'(t) = (dE - dE_z)/dn = \Delta e - c_z/(v \cdot N/L_x) \qquad (3)$$

For the SDF method, $\Delta e$ is normally considered as $P_{ext}$ [18,20,49] and thus $\Delta e'(t)$ can also be treated as $P_{true}(t)$, which then can be further used to calculate $M_{true}$.

From the internal stress data in Fig. 9(a), one can observe that the normal stress fluctuates around 0 GPa while the internal shear stress $\tau_{xz}$ roughly keeps a positive value and gradually declines when adopting the boundary condition of a fully fixed top surface. This means that the normal stress leaves no influence on GB migration, but one part of d$E$ may be used to overcome the impeding effect of $\tau_{xz}$ on migration, which can be quantitatively described in the form of shear strain energy $E_{ss}$ = $0.5V \cdot \tau_{xz}^2/G$. $V$ and $G$ stand for the volume and shear modulus for bicrystal system, respectively. However, $E_{ss}$ is essentially a type of elastic energy and will be dynamically stored and released with the continuous migration of the GB, as supported by Fig. 9(b) which indicates that this energy increment d$E_{ss}$ also fluctuates around 0 eV (i.e., the long-time average of d$E_{ss}$ equaling zero). Therefore, $E_{ss}$ should make no contribution to the overall work-energy relation in the system and the potential influence of $\tau_{xz}$ on GB migration does not need to be considered. Accordingly, Eq. (3) still holds for extracting $P_{true}$.

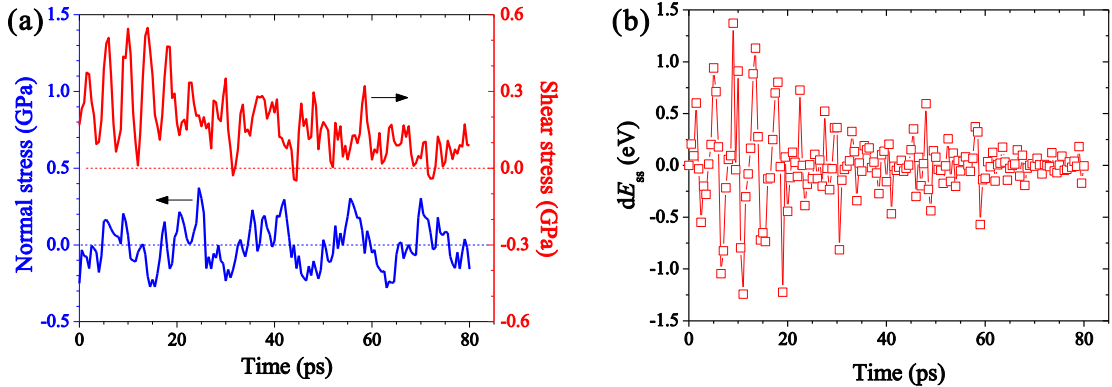

**Fig. 9** Variation of (a) the internal stress and (b) d$E_{ss}$ with the migration time for P1 GB simulated in Fig. 8(b)

Table 2 presents the calculated $P_{true}$ and $M_{true}$ for P1 GB at 500 K according to Eq. (3). In contrast to the early migration stage, the velocity, driving force and mobility at the later stage (e.g., $t > 50$ ps for $L_x$ and $t > 400$ ps for $8L_x$) all only rise slightly and are nearly consistent under $L_x$ and $8L_x$ systems. If we calculate the average value for $v$, $P_{true}$ and $M_{true}$ at the later stage, we can get the average $v$, $P_{true}$ and $M_{true}$ for $L_x$ system as 0.163 nm/ps, 0.0486 eV and 3.341 nm/(ps eV) while $v = 0.162$ nm/ps, $P_{true} = 0.0485$ eV and $M_{true} = 3.353$ nm/(ps eV) for $8L_x$. The relative differences of these three data between the two systems are all lower than 1%. These results and comparisons again emphasize that the acceleration and size-effect in the GB migration velocity have been nearly eliminated through applying one special boundary condition. More importantly, they also signify that we can obtain consistent true mobility values for systems with distinct vertical sizes if further considering the correction of true driving force.

**Table 2** True driving forces and mobility values extracted based on Eq. (3) for P1 GB, simulated at 500 K and 0.06 eV when setting the bottom surface as free while the top one as fully fixed. Both $c_z$ and $v$ were obtained by the least-square fitting into discrete $\Delta E_k$ and $S$ data.

| $L_x$ | | | | $8L_x$ | | | |
|---|---|---|---|---|---|---|---|
| $t$ (ps) | $v$ (nm ps$^{-1}$) | $P_{true}$ (eV) | $M_{true}$ (nm ps$^{-1}$ eV$^{-1}$) | $t$ (ps) | $v$ (nm ps$^{-1}$) | $P_{true}$ (eV) | $M_{true}$ (nm ps$^{-1}$ eV$^{-1}$) |
| 5* | 0.030 | -0.0016 | -18.906 | 40* | 0.035 | 0.0068 | 5.187 |
| 10 | 0.061 | 0.0299 | 2.054 | 80 | 0.065 | 0.0313 | 2.090 |
| 15 | 0.086 | 0.0386 | 2.237 | 120 | 0.089 | 0.0390 | 2.292 |
| 20 | 0.106 | 0.0426 | 2.489 | 160 | 0.108 | 0.0427 | 2.538 |
| 25 | 0.121 | 0.0447 | 2.707 | 200 | 0.123 | 0.0447 | 2.747 |
| 30 | 0.133 | 0.0461 | 2.879 | 240 | 0.134 | 0.0460 | 2.914 |
| 35 | 0.141 | 0.0469 | 3.010 | 280 | 0.142 | 0.0468 | 3.041 |
| 40 | 0.148 | 0.0475 | 3.109 | 320 | 0.149 | 0.0474 | 3.137 |
| 45 | 0.152 | 0.0479 | 3.182 | 360 | 0.153 | 0.0477 | 3.208 |
| 50 | 0.156 | 0.0481 | 3.237 | 400 | 0.157 | 0.0480 | 3.261 |
| 55 | 0.159 | 0.0483 | 3.279 | 440 | 0.159 | 0.0482 | 3.301 |
| 60 | 0.161 | 0.0485 | 3.314 | 480 | 0.161 | 0.0484 | 3.334 |
| 65 | 0.163 | 0.0486 | 3.345 | 520 | 0.163 | 0.0485 | 3.361 |
| 70 | 0.165 | 0.0488 | 3.375 | 560 | 0.164 | 0.0486 | 3.385 |
| 75 | 0.167 | 0.0489 | 3.405 | 600 | 0.166 | 0.0487 | 3.406 |
| 80 | 0.168 | 0.0490 | 3.435 | 640 | 0.167 | 0.0487 | 3.423 |

\* Extracted $P_{true}$ and $M_{true}$ are erroneous due to the low velocity and kinetic energy at the very early stage of migration

It should be noted that the above principle of correcting the driving force should also be applicable to the shear-coupled migration driven by an external shear stress $\tau_{ext}$ (i.e., applicable to Fig. 2(e) and (f)). As shown in the inset of Fig. 10(a) for $\tau_{ext}$-driven migration, $\Delta E_x$ and $\Delta E_y$ remain unchanged while $\Delta E_z$ overall increases linearly with time, in consistency with the tendency shown for the SDF-driven migration in Fig. 8(b). The corresponding correction equation of $P_{true}$ can be given as:

$$P_{true} = \tau_{ext} - \frac{c_z}{v_z \cdot A_{GB}} \qquad (4)$$

where $v_z$ stands for the shear velocity of GB along the $z$ direction.

To compute the true mobility under an external shear stress, Fig. 10 still chooses the P1 GB as an example and the shear stress has been carefully tuned so that the GB migrated at the same velocity as that under the SDF, as shown in Fig. 10(a). Firstly, in comparison with Fig. 2(e), the acceleration in migration is only significant at the early migration stage and turns into uniform migration at the later stage for the current simulation (see blue curve in Fig. 10(a)). Secondly, the true mobility calculated based on the corrected $P_{true}$ according to Eq. (4) experiences a continuous rise and then becomes nearly stable (see the blue square data in Fig. 10(b)). These results do reveal that the accelerated migration under external shear stress can also be largely alleviated by the boundary condition as utilized in Fig. 8(b) for the SDF-driven migration, and that the principle of correcting $P_{true}$ is also applicable to the case of $\tau_{ext}$-driven migration.

What is more, although the displacements or velocities are nearly the same under the SDF and shear stress, the corresponding mobility values under the two types of driving force are significantly different (see Fig. 10(b)). $M_{true}$ by SDF is on average 57.6% lower than that by $\tau_{ext}$. According to the theory recently proposed by Chen et al. [18], the mobility values extracted by applying the driving forces perpendicular and parallel to GB plane can be unified in a mobility tensor as following:

$$\begin{pmatrix} v_x \\ v_y \\ v_z \end{pmatrix} = \begin{pmatrix} M_{xx} & M_{xy} & M_{xz} \\ M_{yx} & M_{yy} & M_{yz} \\ M_{zx} & M_{zy} & M_{zz} \end{pmatrix} \begin{pmatrix} \varphi \\ \tau_y \\ \tau_z \end{pmatrix} \qquad (5)$$

Here $x$ is perpendicular to GB plane while $y$ and $z$ are parallel to GB plane, $\varphi$ is the driving force applied along GB normal (e.g., the synthetic driving force [21,49]), and $\tau_y$ and $\tau_z$ are shear stresses. Moreover, the GB mobility tensor should be symmetric according to the Onsager relation [26], i.e., $M_{xz} = M_{zx}$.

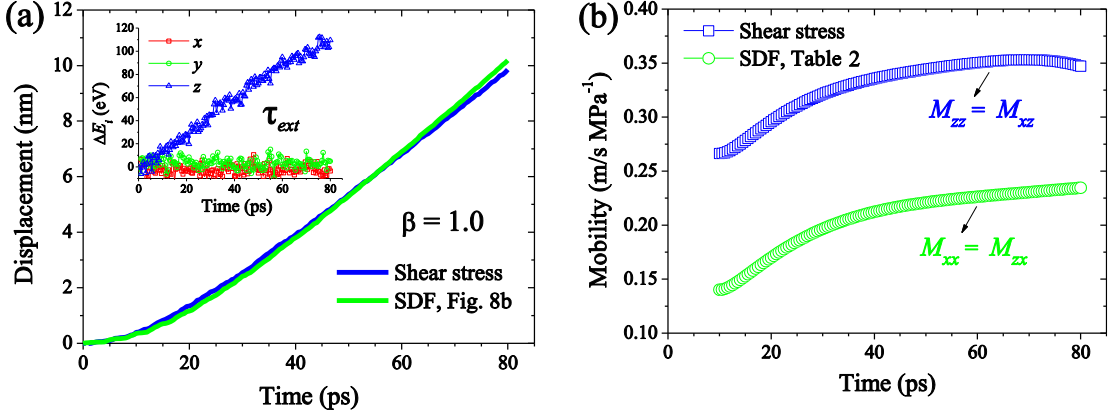

**Fig. 10** Migration results for P1 GB at 500 K when driven under $\tau_{ext}$: (a) displacement along GB normal; (b) true mobility calculated based on the corrected $P_{true}$. The inset in (a) presents the variation of kinetic energy $\Delta E_i$ ($i = x, y, z$). To provide a better comparison of true mobility values between SDF-driven and $\tau_{ext}$-driven migration, the magnitude of $\tau_{ext}$ was chosen deliberately to ensure the corresponding displacement data as close to that by SDF in Fig. 8(b) as possible. The shear stress was applied by adding a shear force 0.0037 eV/Å to a slab of atoms (1 nm thickness) near the bottom surface along $z$ direction, while the top surface was set as fully fixed.

As shown in Fig. 10(a), $\beta$ equals 1.0 under both SDF and shear stress, indicating that $v_x = v_z$ holds under both types of driving force. In such case, we can deduce $M_{xx} = M_{zx}$ for SDF and $M_{zz} = M_{xz}$ for shear stress. Nevertheless, the mobility data presented in Fig. 10(b) reveals $M_{zz} \approx 1.6 M_{xx}$, i.e., $M_{xz} \approx 1.6 M_{zx}$. Therefore, the symmetry of mobility tensor does not hold in the present study, seemingly contradicting to the conclusion in Ref. [26]. Recently, we have systematically investigated the GB mobility tensor and the effects of temperature and external driving force on its symmetry based on atomistic simulations of force-driven and force-free migration for the twist Ni Σ15 (2 1 1) (P14) GB [52]. It is found that the symmetry holds at low driving force limit while fails at high driving forces, i.e., the symmetry strongly depending on the magnitude of driving force. Therefore, the non-equal $M_{xz}$ and $M_{zx}$ as shown in Fig. 10(b) is due to the relatively large driving forces that have been applied in the current study, which is also consistent with the previous studies that large driving forces can significantly change the underlying mechanisms for GB migration [20].

## 4. Conclusions

In this study, we carried out atomistic simulations to investigate the migration behaviors of several grain boundaries (GBs) in Ni, recognizing their significant impact on thermal and mechanical responses in polycrystalline materials. Our findings have led to several important conclusions:

(1) The migration displacements of some GBs driven under a constant external force do not follow the widely assumed linear relation with migration time. Instead, the corresponding velocity gradually increases as the boundary migration progresses, and it is negatively correlated with the cell size perpendicular to the GB plane, regardless of the driving force magnitude and type. These observations highlight the need to calibrate previously published migration results for such GBs, as neglecting the acceleration and size effect may lead to inaccurate conclusions.

(2) The acceleration and vertical size effect in migration are unique to shear-coupling GBs, which exhibit a rise in the kinetic energy component along the shear direction. This rise in kinetic energy results in the true driving force for GB migration being lower than the applied value but continuing to increase, leading to accelerated migration. However, at higher temperatures, the acceleration transforms into uniform migration, and the size effect diminishes accordingly.

(3) Among various attempts to eliminate or alleviate the acceleration and size-dependency in migration, we found that setting the cell surface in the forward direction of GB migration as fully fixed while keeping the surface in the backward direction free is a simple yet effective approach. This boundary condition applies to both shear stress and synthetic driving force (SDF)-driven migration and preserves the shear-coupling behaviors and migration mechanism.

(4) Through a quantitative analysis of the work-energy relation in the bicrystal system, we can determine the true driving force for GBs exhibiting accelerated migration. By adopting the specific boundary condition and correcting the true driving force, we obtain consistent true mobility values for GBs with distinct vertical sizes. Furthermore, our calculations suggest that the symmetry of the mobility tensor may break down under excessively large driving forces.

In conclusion, this study sheds light on the complex behaviors of GB migration and provides valuable insights into manipulating migration phenomena through boundary conditions and internal stress. The proposed method for extracting the true driving force and true mobility offers a robust approach for future simulations of GB migration-related phenomena, such as crack propagation, recrystallization, and grain growth, contributing to the advancement of materials science and engineering.

## Acknowledgment

The authors thank Dr. David L Olmsted for sharing the 388 Ni GB structure database. This research was supported by the National Natural Science Foundation of China (Grant No. 52065045) and NSERC Discovery Grant (RGPIN-2019-05834), Canada, and the use of computing resources provided by Compute/Calcul Canada.

**References:**

[1] G. Gottstein, L.S. Shvindlerman, Grain Boundary Migration in metals: thermo-dynamics, kinetics, Applications, CRC press, 2009.

[2] M.A. Meyers, A. Mishra, D.J. Benson, Mechanical properties of nanocrystalline materials, Prog. Mater. Sci. 51(2006) 427–556.

[3] Z. Shan, E.A. Stach, J.M.K. Wiezorek, J.A. Knapp, D.M. Follstaedt, S.X. Mao, Grain boundary-mediated plasticity in nanocrystalline nickel, Science 305(2004) 654–657.

[4] Q. Zhu, H. Zhou, Y. Chen, G. Cao, C. Deng, Z. Zhang, J. Wang, Atomistic dynamics of disconnection-mediated grain boundary plasticity: A case study of gold nanocrystals, J. Mater. Sci. Technol. 125(2022) 182–191.

[5] J. Hu, Y.N. Shi, X. Sauvage, G. Sha, K. Lu, Grain boundary stability governs hardening and softening in extremely fine nanograined metals, Science 355(2017) 1292–1296.

[6] T.H. Fang, N.R. Tao, K. Lu, Tension-induced softening and hardening in gradient nanograined surface layer in copper, Scr. Mater. 77(2014) 17–20.

[7] D.S. Gianola, B.G. Mendis, X.M. Cheng, K.J. Hemker, Grain-size stabilization by impurities and effect on stress-coupled grain growth in nanocrystalline Al thin films, Mater. Sci. Eng. A 483(2008) 637–640.

[8] J. Hu, J.X. Li, Y.-N. Shi, Suppression of grain boundary migration at cryogenic temperature in an extremely fine nanograined Ni-Mo alloy, J. Mater. Sci. Technol. 57(2020) 65–69.

[9] M. Upmanyu, D.J. Srolovitz, L.S. Shvindlerman, G. Gottstein, Misorientation dependence of intrinsic grain boundary mobility: simulation and experiment, Acta Mater. 47 (1999) 3901–3914.

[10] J.E. Brandenburg, D.A. Molodov, On shear coupled migration of low angle grain boundaries, Scr. Mater. 163(2019) 96–100.

[11] Q. Zhu, G. Cao, J. Wang, C. Deng, J. Li, Z. Zhang, S.X. Mao, In situ atomistic observation of

disconnection-mediated grain boundary migration, Nat. Commun. 10 (2019) 156.

[12] H. Zhang, D.J. Srolovitz, Simulation and analysis of the migration mechanism of Σ5 tilt grain boundaries in an fcc metal, Acta Mater. 54 (2006) 623–633.

[13] J.W. Cahn, Y. Mishin, A. Suzuki, Coupling grain boundary motion to shear deformation, Acta Mater. 54 (2006) 4953–4975.

[14] J. Han, S.L. Thomas, D.J. Srolovitz, Grain-boundary kinetics: A unified approach, Prog. Mater. Sci. 98 (2018) 386–476.

[15] E.R. Homer, O.K. Johnson, D. Britton, J.E. Patterson, E.T. Sevy, G.B. Thompson, A classical equation that accounts for observations of non-Arrhenius and cryogenic grain boundary migration, npj Comput. Mater. 8 (2022) 1–9.

[16] T. Yu, S. Yang, C. Deng, Survey of grain boundary migration and thermal behavior in Ni at low homologous temperatures, Acta Mater. 177 (2019) 151–159.

[17] D.L. Olmsted, S.M. Foiles, E.A. Holm, Grain boundary interface roughening transition and its effect on grain boundary mobility for non-faceting boundaries, Scr. Mater. 57 (2007) 1161–1164.

[18] D.L. Olmsted, E.A. Holm, S.M. Foiles, Survey of computed grain boundary properties in face-centered cubic metals—II: Grain boundary mobility, Acta Mater. 57 (2009) 3704–3713.

[19] M.I. Mendelev, C. Deng, C.A. Schuh, D.J. Srolovitz, Comparison of molecular dynamics simulation methods for the study of grain boundary migration, Model. Simul. Mater. Sci. Eng. 21 (2013) 045017.

[20] C. Deng, C.A. Schuh, Diffusive-to-ballistic transition in grain boundary motion studied by atomistic simulations, Phys. Rev. B. 84 (2011) 214102.

[21] L. Yang, S.Y. Li, A modified synthetic driving force method for molecular dynamics simulation of grain boundary migration, Acta Mater. 100 (2015) 107–117.

[22] M.J. Rahman, H.S. Zurob, J.J. Hoyt, A comprehensive molecular dynamics study of low-angle grain boundary mobility in a pure aluminum system, Acta Mater. 74 (2014) 39–48.

[23] B. Lin, K. Wang, F. Liu, Y. Zhou, An intrinsic correlation between driving force and energy barrier upon grain boundary migration, J. Mater. Sci. Technol. 34(2018) 1359–1363.

[24] Z.T. Trautt, M. Upmanyu, A. Karma, Interface mobility from interface random walk, Science 314 (2006) 632–635.

[25] K. Chen, J. Han, S. L. Thomas, D. J. Srolovitz, Grain boundary shear coupling is not a grain boundary property, Acta Mater. 167 (2019) 241–247.

[26] K. Chen, J. Han, X. Pan, D.J. Srolovitz, The grain boundary mobility tensor, Proc. Natl. Acad. Sci. 117 (2020) 4533–4538.

[27] L. Zhou, H. Zhang, D.J. Srolovitz, A size effect in grain boundary migration: A molecular dynamics study of bicrystal thin films, Acta Mater. 53 (2005) 5273–5279.

[28] J. Humberson, E.A. Holm, Anti-thermal mobility in the Σ3 [111] 60°{11 8 5} grain boundary in nickel: mechanism and computational considerations, Scr. Mater. 130 (2017) 1–6.

[29] C.P. Race, J. von Pezold, J. Neugebauer, Role of the mesoscale in migration kinetics of flat grain boundaries, Phys. Rev. B 899 (2014) 214110.

[30] A. Rajabzadeh, F. Mompiou, M. Legros, N. Combe, Elementary mechanisms of shear-coupled grain boundary migration, Phys. Rev. Lett. 110 (2013) 265507.

[31] N. Combe, F. Mompiou, M. Legros, Disconnections kinks and competing modes in shear-coupled grain boundary migration, Phys. Rev. B. 93 (2016) 024109.

[32] Y. Deng, C. Deng, Size and rate dependent grain boundary motion mediated by disconnection nucleation, Acta Mater. 131 (2017) 400–409.

[33] T. Gorkaya, D.A. Molodov, G. Gottstein, Stress-driven migration of symmetrical <100> tilt grain boundaries in Al bicrystals, Acta Mater. 57(2009) 5396–5405.

[34] G. Gottstein, L.S. Shvindlerman, The compensation effect in thermally activated interface processes, Interface Sci. 6 (1998) 267–278.

[35] H. Hahn, H. Gleiter, The effect of pressure on grain growth and boundary mobility, Scr. Metall. 13(1979) 3–6.

[36] D.A. Molodov, B.B. Straumal, L.S. Shvindlerman, The effect of pressure on migration of <001> tilt grain boundaries in tin bicrystals, Scr. Metall. 18(1984) 207–211.

[37] J.W. Cahn, The impurity-drag effect in grain boundary motion, Acta Metall. 10(1962) 789–798.

[38] G. Gottstein, L.S. Shvindlerman, On the orientation dependence of grain boundary migration, Scr. Metall. Mater. 27 (1992) 1515–1520.

[39] F. Ulomek, V. Mohles, Separating grain boundary migration mechanisms in molecular dynamics simulations, Acta Mater. 103 (2016) 424–432.

[40] H. Zhang, M.I. Mendelev, D.J. Srolovitz, Computer simulation of the elastically driven migration of a flat grain boundary, Acta Mater. 52 (2004) 2569–2576.

[41] D. Farkas, S. Mohanty, J. Monk, Linear grain growth kinetics and rotation in nanocrystalline Ni, Phys. Rev. Lett. 98 (2007) 165502.

[42] L. Yang, C. Lai, S.Y. Li, A survey of the crystallography-dependency of twist grain boundary mobility in Al based on atomistic simulations, Mater. Lett. 263 (2020) 127293.

[43] S. Plimpton, Fast parallel algorithms for short-range molecular dynamics, J. Comput. Phys. 117 (1995) 1–19.

[44] S.M. Foiles, J.J. Hoyt, Computation of grain boundary stiffness and mobility from boundary fluctuations, Acta Mater. 54 (2006) 3351–3357.

[45] A. Stukowski, Visualization and analysis of atomistic simulation data with ovito–the open visualization tool, Model. Simul. Mater. Sci. Eng. 18 (2009) 015012.

[46] S.P. Coleman, D.E. Spearot, S.M. Foiles, The effect of synthetic driving force on the atomic mechanisms associated with grain boundary motion below the interface roughening temperature, Comput. Mater. Sci. 86 (2014) 38–42.

[47] E.R. Homer, S.M. Foiles, E.A. Holm, D.L. Olmsted, Phenomenology of shear-coupled grain

boundary motion in symmetric tilt and general grain boundaries, Acta Mater. 61 (2013) 1048–1060.

[48] C. Deng, C.A. Schuh, Atomistic simulation of slow grain boundary motion, Phys. Rev. Lett. 106 (2011) 045503.

[49] A.A. Schratt, V. Mohles, Efficient calculation of the ECO driving force for atomistic simulations of grain boundary motion, Comput. Mater. Sci. 182 (2020) 109774.

[50] S. Starikov, A. Abbass, R. Drautz, M. Mrovec, Disordering complexion transition of grain boundaries in bcc metals: Insights from atomistic simulations, Acta Mater. 261 (2023) 119399.

[51] S.L. Thomas, K. Chen, J. Han, P.K. Purohit, D.J. Srolovitz, Reconciling grain growth and shear-coupled grain boundary migration, Nat. Commun. 8 (2017) 1764.

[52] X.Y. Song, L. Yang, C. Deng, Computing the intrinsic grain boundary mobility tensor, https://doi.org/10.48550/arXiv.2212.11462.

**Supplemental Material:**

**Unusual acceleration and size effects in grain boundary migration with shear coupling**

Liang Yang[a], Xinyuan Song[b], Tingting Yu[c], Dahai Liu[a*], Chuang Deng[b,*]

[a] School of Aeronautical Manufacturing Engineering, Nanchang Hangkong University, Nanchang 330063, China

[b] Department of Mechanical Engineering, University of Manitoba, Winnipeg, MB R3T 2N2, Canada

[c] School of Aviation and Mechanical Engineering, Changzhou Institute of Technology, Changzhou, Jiangsu 213032, China.

* Corresponding author: dhliu@nchu.edu.cn(D. Liu), Chuang.Deng@umanitoba.ca (C. Deng)

## 1. Supporting results for the transition of acceleration migration into uniform migration

In Fig. 1 of the main text, we can readily observe the accelerated migration and negative dependency of velocity on the vertical cell size for P1 GB at 500 K. Nevertheless, when raising the temperature to 1000 K, both the acceleration and size effect disappear though the shear coupling still exists. The results in Fig. s1 for P148 GB suggest that the transition from accelerated into uniform migration, induced by the temperature, does also exist for other shear-coupling GBs, regardless of the magnitude of driving force.

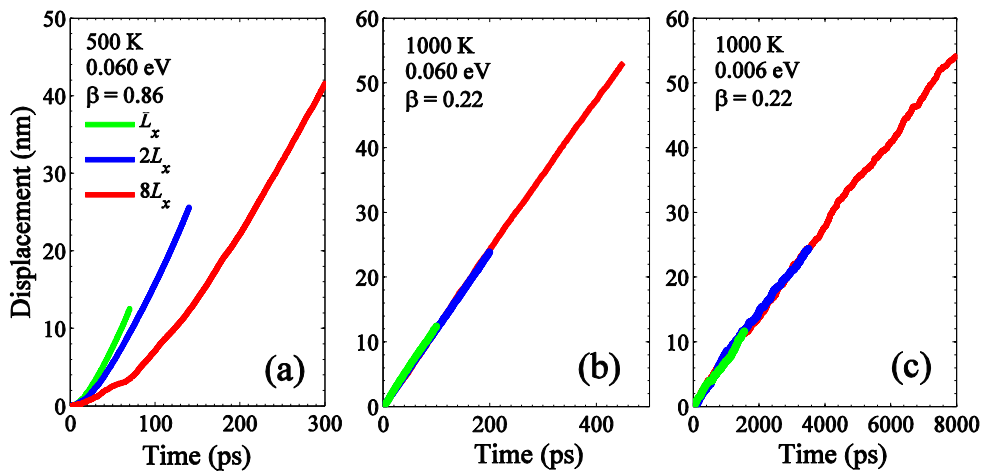

**Fig. s1** Variation of GB displacement with the vertical size for P148 GB, simulated under distinct temperatures and driving forces. Concerning simulation settings were the same as Fig. 1 in the main text.

## 2. Supporting results for the effective alleviation of acceleration in migration

When adopting one special kind of boundary condition for GBs exhibiting acceleration, i.e., setting the bottom surface as free while the top one as fully fixed, the acceleration will only be significant at the early migration stage and gradually turn into uniform migration at the later stage. In such case, both the acceleration in migration and size effect can be effectively alleviated. Nevertheless, we only present the complete displacement and kinetic energy data for $L_x$ system due to the limited scope of the main text. Here Fig. s2 shows the complete data for all four distinct systems considered in the main text to support the above conclusion.

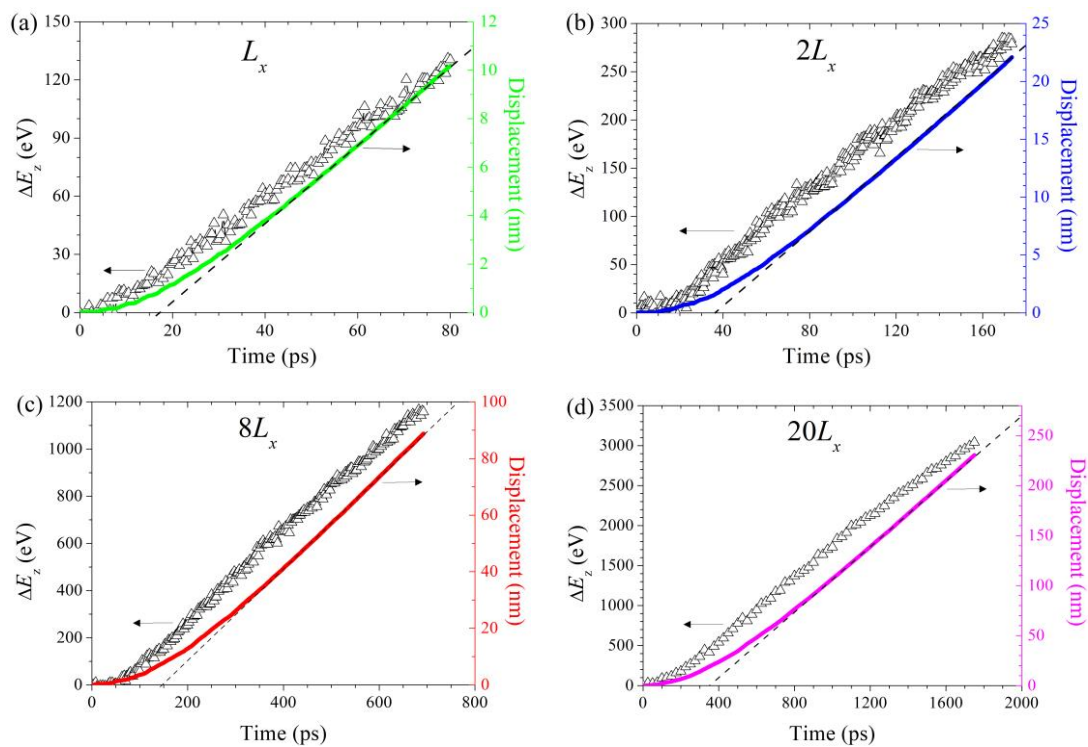

**Fig. s2** GB displacement and shear kinetic energy data for P1 GB with different vertical sizes, simulated under 500 K and 0.06 eV when setting the bottom surface as free while the top one as fully fixed.

## 3. Supporting results for the effect of boundary conditions on GB migration

When attempting to alleviate the acceleration in migration and size effect, we have performed simulations adopting two free boundaries or setting one boundary as free while the other as fixed for P1 GB. The corresponding GB displacement *vs.* time date under these boundary conditions have been given in Fig. 8(a) in the main text. To more intuitively exhibit the differences in the tendencies of migration velocity with the time under various boundary conditions, Fig. s3 directly presents the instant velocity *vs.* time data. It is clear that acceleration is only significant at the early migration stage and gradually turns into uniform migration at the later stage when the top surface was fully fixed, while the migration velocities under the other boundary conditions continue to rise during the whole migration process.

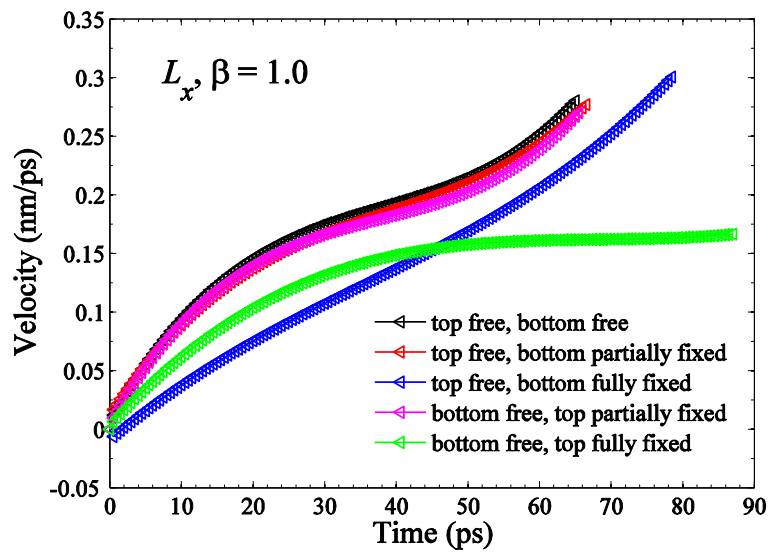

**Fig. s3** Comparison of GB migration velocity *vs.* time data under various boundary conditions, simulated for P1 GB at 500 K and 0.06 eV. The corresponding displacement data can be found in Fig. 8(a) in the main text.